\documentstyle[epsf,aps]{revtex}
\renewcommand{\thefootnote}{\fnsymbol{footnote}}
\begin{document}
\begin{flushright}

Columbia preprint CU--TP--824\\
Duke preprint DU--TH--141
\end{flushright}
\vspace*{1cm} 
\setcounter{footnote}{1}
\begin{center}
{\Large\bf Classical Gluon Radiation in Ultrarelativistic
Nucleus--Nucleus Collisions}
\\[1cm]
Yuri V.\ Kovchegov $^1$ and Dirk H.\ Rischke $^2$ \\ ~~ \\
{\it $^1$ Department of Physics, Columbia University} \\ 
{\it New York, New York 10027} \\ ~~ \\
{\it $^2$ Department of Physics,  Duke University} \\ 
{\it Box 90305, Durham, North Carolina 27708-0305} \\ ~~ \\ ~~ \\
\end{center}
\begin{abstract} 
 The classical Yang--Mills equations are solved perturbatively in
covariant gauge for a collision of two ultrarelativistic nuclei. The
nuclei are taken as ensembles of classical color charges on eikonal
trajectories. The classical gluon field is computed in coordinate space
up to cubic order in the coupling constant $g$.  We
construct the Feynman diagrams corresponding to this field and show
the equivalence of the classical and diagrammatic approaches. An
argument is given which demonstrates that at higher orders in $g$
the classical description of the process breaks down. 
As an application, we calculate the energy, number, and multiplicity 
distributions of produced soft gluons and reproduce earlier
results by Gunion and Bertsch and by Kovner, McLerran, and Weigert.

\ \\ PACS number(s): 12.38.Bx, 12.38.Aw, 24.85.+p, 25.75-q

\end{abstract}
\renewcommand{\thefootnote}{\arabic{footnote}}
\setcounter{footnote}{0}

\section{Introduction}

  Ultrarelativistic heavy-ion collisions ($\sqrt{s} \sim 200$ AGeV) at
the Relativistic Heavy Ion Collider (RHIC) at Brookhaven aim towards
an understanding of the properties of nuclear matter under extreme
conditions \cite{harrismuller}.  It was argued that the
extraordinarily high energy and particle number densities reached in
central nuclear collisions at RHIC, $\epsilon \sim 10 - 20\, {\rm
GeV\, fm}^{-3}$, $dN/dy \sim 1000$ \cite{miklosQM}, could lead to
rapid (local) thermalization of matter \cite{hotglue} and thus to the
creation of the so-called quark--gluon plasma (QGP) \cite{muller}, a
state predicted by finite temperature lattice QCD calculations
\cite{lattice}, where chiral symmetry is restored and quarks and
gluons are deconfined.

  In order to assess whether this state can actually be formed in an
ultrarelativistic nuclear collision one has to gain a better
understanding of the initial conditions and, at these energies
predominantly hard, parton--parton scattering processes in the early
stage preceding (local) thermodynamical equilibrium. While event
generators based on individual parton--parton scattering processes
\cite{klaus} have been developed, their respective predictions for the
range of accessible energy and particle number densities differ
widely.

  One of the main reasons is the poor understanding of the initial
conditions for the nuclear reaction. Recently, McLerran and
Venugopalan have made considerable progress in a classical approach
\cite{Larry} to construct the gluon field at small values of
$x$. Their treatment is somewhat similar to the approach used by
Mueller for constructing the wave function and gluon structure
function of a heavy quarkonium state \cite{Mueller}. At small $x$, 
the nucleonic structure is dominated by gluons, and thus a proper
description of gluon dynamics in this kinematic region 
is vital for understanding the initial conditions and the subsequent 
pre-equilibrium stage in nuclear collisions.

  The McLerran--Venugopalan model \cite{Larry} considers a very large
nucleus moving at ultrarelativistic velocity, which consequently
appears in the laboratory frame as a ``pancake'' in the transverse
plane.  It is assumed \cite{Larry} that due to the large size of the
nucleus the (transverse, two--dimensional) color charge density
$\rho(\underline{x})$ is large (i.e., in a higher-dimensional
representation of the color algebra) so that in a certain kinematic
region the soft gluon field produced by these color charges is
effectively {\em classical}, and can thus be obtained by solving the
{\em classical\/} Yang--Mills equations of motion. This field can then
be used to compute the gluon distribution function. Quantum effects
can be implemented as corrections to the classical field.

  The kinematic region for which this approximation is valid is given
by the following consideration \cite{Larry,Alex}: the strong coupling
constant $\alpha_S \equiv g^2/4 \pi$ should be small, therefore the
typical gluon transverse momenta in the problem should satisfy
$k_\perp \gg \Lambda_{\rm QCD}$. On the other hand, the gluon transverse
momenta should be sufficiently ``soft'', such that the
gluons do not resolve individual color charges but couple to the
classical color charge density. At very high transverse momenta
quantum effects become important. Therefore, we have to limit ourselves
to the region where $k_\perp \ll \mu$, with $\mu^2$ being the average
color charge density squared. The momentum fraction $x$ of the gluons
should be small enough so that the nucleus appears coherent in the
longitudinal direction.

  To facilitate the inclusion of quantum corrections, the authors
of \cite{Larry} searched for the classical gluon field of such a 
nucleus in the light-cone gauge. The solution of the equations of motion 
requires to take the longitudinal extension of the nucleus into account, 
i.e., one must not take the nucleus to be infinitely thin in the longitudinal
direction, as was assumed originally in \cite{Larry}.  The classical
field of a single ultrarelativistic nucleus is the non-Abelian
Weizs\"{a}cker--Williams field. It was computed in
\cite{Jamal,yuri}. In the approach pursued in \cite{yuri} the
nucleus was assumed to be an ensemble of nucleons consisting of
point-like (valence) color charges. Instead of a smooth
two--dimensional color charge density $\rho(\underline{x})$, this
quantity is a sum of $\delta$ functions in \cite{yuri}. The limit of
applicability of the classical approximation, as well as the structure
of the non-Abelian Weizs\"{a}cker--Williams field in terms of Feynman
diagrams has been discussed in \cite{yuri'}.  That model also allowed
for an explicit calculation of the average color charge density
squared, $\mu^2$.

One of the goals of the approach of McLerran and Venugopalan is to
obtain the BFKL equation for the structure function of soft gluons,
and, if possible, derive corrections to this equation which account
for nuclear shadowing.  Recently, a first step in this direction has
been made: the BFKL equation was obtained via a renormalization group
approach \cite{Andrei}.

  A collision of two ultrarelativistic nuclei of the type advertised
in \cite{Larry} was considered in \cite{Alex}. The Yang--Mills
equations were simplified assuming that the solution for the gluon
field in the forward light-cone is boost-invariant. The equations were
then solved perturbatively to first order in the corrections to the
Abelian solution in the gauge $x_- A_+ + x_+ A_- = 0$ (where
$x_\pm = (t \pm z)/\sqrt{2}$, and $A_\pm = (A^0 \pm A^z)/\sqrt{2}$
are the light-cone components of the gluon field).
      
\begin{figure}
\begin{center}
\epsfxsize=7cm
\epsfysize=7cm
\leavevmode
\hbox{ \epsffile{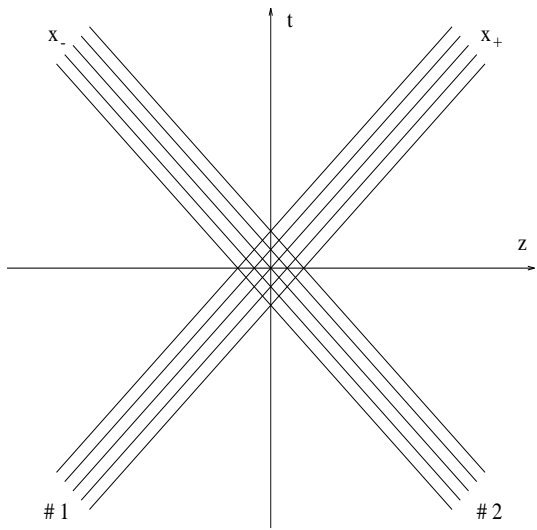}}
\end{center}
\caption{The nuclear collision as envisaged here (for details, see text).}
\label{coll}
\end{figure}

  In this paper, we also focus on an ultrarelativistic nuclear
collision, but we employ the approach of \cite{yuri} to describe the
nuclei.  Each nucleus is moving with the speed of light. The nuclei
are taken to be ensembles of nucleons, consisting of point-like color
charges (valence quarks). The first nucleus is supposed to move in the
``+''--direction, the second in the ``--''--direction, see Fig.\ 1.  
In contrast to \cite{Larry,Alex},
they {\em do\/} have a longitudinal extension, i.e., a color charge in
nucleus 1 has a fixed $x_-$--component, $x_{-i}$, which is different
for each charge, $x_{-i} \neq x_{-k}, \, i \neq k$, and similar for
nucleus 2, where the charges have fixed $x_+$--components, $y_{+j}$,
(of course, all charges have different transverse coordinates
$\underline{x}_i,\, \underline{y}_j$ as well).

  During the collision, we assume the momenta of the charges to remain
unchanged (eikonal trajectories). This is certainly justified, since
we consider the initial momenta of the charges to be rather large (if
not infinite).  The nuclei just pass through each other and continue
their motion along the light-cone (see Fig.\ \ref{coll}).  As in
\cite{Alex}, we also solve the classical Yang--Mills equations
perturbatively to first order in the corrections to the Abelian
solution and obtain the classical, radiated gluon
field. In contrast to \cite{Alex}, however, we shall work {\em
exclusively\/} in the covariant (Lorentz) gauge. The advantage of this
gauge is that, for the case of a single ultrarelativistic nucleus, the
classical gluon field is identical to the solution of the
corresponding Abelian problem \cite{yuri}. Moreover, as will be shown
below, for the collision of two nuclei, one can easily relate the
classical solution to that of a diagrammatic approach in terms of the
usual Feynman rules in covariant gauge [such rules do not exist in
momentum space for the gauge chosen in \cite{Alex}].

  The outline of the paper is as follows. In Section II we discuss the
lowest order solution to the Yang--Mills equations. 
This is just the solution to the corresponding Abelian problem, i.e., 
the fields generated by the color charges in the nuclei are simply superposed
and no real gluons are produced.
In Section III we find the classical field to order $g^3$.
This is the first (and lowest) order correction to the Abelian
solution.  We shall derive an explicit expression for the field in
coordinate space. We establish the correspondence between this
classical result and a particular set of Feynman diagrams. This proves
that, at this order, the classical field is the (major) source of soft
gluon production. We shall argue that at higher orders the classical
description will fail, since already at order $g^5$ non-classical
contributions to the gluon field become important.  In Section IV we
calculate energy, number, and multiplicity distributions of the
produced gluons. As expected, their form is similar to the one
found previously in \cite{Alex,gunion}. However, the prefactor of our
result is different from that in \cite{Alex}, while it agrees with the
result of \cite{gunion} and \cite{Miklos}.

Our units are $\hbar = c =1$, and the metric tensor is $g^{\mu \nu} =
{\rm diag} (+,-,-,-)$. Light-cone coordinates are defined in the usual
way, $a_\pm \equiv (a^0 \pm a^z)/\sqrt{2},\, \partial_\mp \equiv
\partial / \partial x_\pm$. The notation for transverse vectors is
$\underline{a} = (a^x,a^y)$.

\section{Classical solution to lowest order in the coupling constant}

     We consider two nuclei with mass numbers $A_1, \, A_2$ moving
towards each other with ultrarelativistic velocities, $v_{1,2} \simeq
\pm 1$, along the $z$--axis (cf.\ Fig.\ \ref{coll}).  The nuclei are
taken as ensembles of nucleons \cite{yuri}.  In order to simplify the
color algebra each ``nucleon'' consists of a quark--antiquark
pair. These valence quarks and antiquarks are confined inside the
nucleons (visualized as spheres of equal radius in the rest frame of
each nucleus).  In order to construct the solution, nucleons inside
the nucleus and valence charges inside the nucleons are assumed to be
``frozen'', i.e., they have definite light-cone (and transverse)
coordinates which, due to our assumption of eikonal trajectories for
the individual charges, will not change throughout the calculation. We
label the coordinates of the quarks in nucleus 1 by $x_{-i},\,
\underline{x}_i$, $i=1, \ldots ,A_1$, and those of nucleus 2 by
$y_{+j},\, \underline{y}_j$, $j=1, \ldots ,A_2$.  Antiquark
coordinates follow this notation with an additional prime.  As in
\cite{yuri}, the nuclei are supposed to be sufficiently ``dilute'',
such that the distance between the nucleons is much larger than the
nucleon's size.

The goal of this section is to solve the classical QCD equations of motion,
\begin{equation} 
D_\mu F^{\mu \nu} = J^{\nu}\,\, ,
\label{eom}
\end{equation}
to lowest order in the strong coupling constant (our convention for
the covariant derivative is $D_\mu \equiv \partial_\mu -ig 
[A_\mu,\, \cdot\, ]$). 
We shall work exclusively in the covariant gauge ($\partial_\mu A^\mu = 0$).  
In this gauge, the equations of motion (\ref{eom}) can be cast
into the form
\begin{equation} 
\Box A^{\mu} = J^\mu + ig \, [A_\nu , \partial^\nu A^\mu + F^{\nu \mu}
]\,\, ,
\label{eom2}
\end{equation}
where $\Box$ is the d'Alembertian operator. In this form, it is easy
to solve the equations perturbatively, as will be outlined in the
following.

   To lowest order in $g$, i.e., order $g$, the commutator term
on the right-hand side of Eq.\ (\ref{eom2}) does not contribute, since
the field itself is already of this order [see Eq.\ (\ref{clsol1}) below].  
To this order and in covariant gauge, the classical current can be taken 
as a sum of the currents for each individual nucleus, as given in \cite{yuri}:
\begin{mathletters} \label{current1}
\begin{eqnarray} 
J_+^{(1)} & = & g \sum_{i=1}^{A_1} T^a \, (T_i^a)\, [\delta(x_- -
x_{-i})\, \delta( {\underline x}-{\underline x}_i ) - \delta(x_- -
x'_{-i})\, \delta( {\underline x}-{\underline x}'_i ) ]\,\, , \\ 
J_-^{(1)}
& = & g \sum_{j=1}^{A_2} T^a \, (\tilde{T}_j^a)\, [\delta(x_+ -
y_{+j})\, \delta( {\underline x}-{\underline y}_j ) - \delta(x_+ -
y'_{+j})\, \delta( {\underline x}-{\underline y}'_j )]\,\, , \\ 
{\underline J}^{(1)} & = & 0\,\, ,
\end{eqnarray}
\end{mathletters}
where $T^a$ are the generators of $SU(N_c)$, and $(T_i^a)$ and
$(\tilde{T}_j^a)$ are color matrices which represent the color charge
of the quark in the color space of nucleon $i$ in the first and
nucleon $j$ in the second nucleus, respectively [see \cite{yuri}]. The
current (\ref{current1}) takes into account that antiquarks have the
opposite color charge and thus ensures color neutrality for each
nucleon.

  The classical gluon field satisfying the Yang--Mills equations for a
{\em single\/} ultrarelativistic nucleus of our type was found in
\cite{yuri}. To lowest order in the coupling constant the solution of
Eq.\ (\ref{eom2}) for two nuclei will be just a sum of the
solutions for single nuclei, since the equations of motion are Abelian
($\Box A_{\mu}^{(1)} = J_\mu^{(1)}$), and thus linear.  Therefore, as
one readily confirms by an explicit calculation, the solution of the
Yang--Mills equations to order $g$ is \cite{yuri}:
\begin{mathletters}\label{clsol1}
\begin{eqnarray} 
A_{+}^{(1)} & = & {-} { g \over {2 \pi }} \sum_{i=1}^{A_1} T^a \,
 (T_i^a)\, \left[ \delta(x_{-}-x_{-i}) \, \ln
 (|{\underline{x}}-{\underline{x}}_i| \lambda ) -
 \delta(x_{-}-x'_{-i}) \, \ln (|{\underline{x}}-{\underline{x}}'_i|
 \lambda ) \right]\,\, , \\ 
A_{-}^{(1)} & = & {-} { g \over {2 \pi }}
 \sum_{j=1}^{A_2} T^a\, (\tilde{T}_j^a) \, \left[ \delta(x_{+}-y_{+j})
 \, \ln (|{\underline{x}}-{\underline{y}}_j| \lambda )
 -\delta(x_{+}-y'_{+j})\, \ln (|{\underline{x}}-{\underline{y}}'_j|
 \lambda ) \right]\,\, , \\ 
{\underline A}^{(1)} & = & 0\,\, ,
\end{eqnarray} 
\end{mathletters}
where $\lambda$ enters as an infrared cut-off. In a sense, it acts as
a gauge parameter that sets the scale of the gauge potential. The
associated field strength tensor is independent of $\lambda$,
\begin{mathletters}
\begin{eqnarray}
F_{+-}^{(1)} & = & 0 \,\,\, , \\ 
F_{+\perp}^{(1)} & = & { g \over {2 \pi}} 
\sum_{i=1}^{A_1} T^a \, (T_i^a) \, \left(
\delta(x_{-}-x_{-i})\, {{\underline{x}}-{\underline{x}}_i \over
|{\underline{x}}-{\underline{x}}_i|^2}
-\delta(x_{-}-x'_{-i})\, {{\underline{x}}-{\underline{x}'}_i \over
|{\underline{x}}-{\underline{x}'}_i|^2} \right) \,\, , \\ 
F_{-\perp}^{(1)} & = & { g \over {2 \pi}} \sum_{j=1}^{A_2} 
T^a\, (\tilde{T}_j^a) \, \left(
\delta(x_{+}-y_{+j})\, {{\underline{x}}-{\underline{y}}_j \over
|{\underline{x}}-{\underline{y}}_j|^2}
-\delta(x_{+}-y'_{+j})\, {{\underline{x}}-{\underline{y}'}_j \over
|{\underline{x}}-{\underline{y}'}_j|^2} \right)\,\, , \\ 
F_{i j}^{(1)} & = & 0
\,\, , \,\,\,\, i,j=x,y \,\,\, .
\end{eqnarray}
\end{mathletters}
We note that the field strength is zero in the forward light-cone.
This is of course reasonable, because to order $g$ there are no
interactions between the point-like charges constituting the
nuclei. Therefore, no real gluons are produced.

\section{Gluon field to next-to-lowest order}

\subsection{Formal solution of the equations of motion}

   In this section we compute the solution of the Yang--Mills
equations to order $g^3$. As we shall see, the equations of motion
(\ref{eom2}) are linear to each order, and we therefore focus first,
for the sake of simplicity, on the case of a collision of two {\em
single\/} point-like color charges, for instance the quark from
nucleon $i$ in nucleus 1 and the quark from nucleon $j$ in nucleus 2.
The generalization to the nuclear collision is then straightforward:
the solution is a simple superposition of the solutions emerging from
each individual collision (i.e., a sum over $i$ and $j$ and over the
respective quark--quark, quark--antiquark, antiquark--quark, and
antiquark--antiquark scatterings).

   To order $g^3$ the equations of motion (\ref{eom2}) read:
\begin{equation} 
\Box A_{\mu}^{(3)} = J_\mu^{(3)} + ig \, [A^{(1)\nu} ,
\partial_\nu A_\mu^{(1)} + F_{\nu \mu}^{(1)} ]\,\,\, ,
\label{eom3}
\end{equation}
where $A_{\mu}^{(3)}$ and $J_\mu^{(3)}$ are the
contributions to the gluon field and the fermionic current to order $g^3$.
In order to solve these equations we have to first
determine $J_\mu^{(3)}$.  The most simple approach is to exploit
(non-Abelian) current conservation $D_\mu \,J^\mu=0$. One obtains:
\begin{eqnarray}
\partial_\mu J^{(3)\mu} & = & ig\, [A_+^{(1)},J_-^{(1)}] + ig\,
[A_-^{(1)},J_+^{(1)}] = {g^3 \over {2 \pi}} \, f^{abc}\, T^a \,
(T_i^b)\, (\tilde{T}_j^c) \nonumber \\ & & \times \,\,\, \delta (x_- -
x_{-i})\, \delta (x_+ - y_{+j}) \, \left[ \delta({\underline
x}-{\underline y}_j) - \delta({\underline x}-{\underline x}_i) \right]
\, \ln (|{\underline x}_i - {\underline y}_j| \lambda) \, \,\,\, .
\label{cons}
\end{eqnarray}
The charges are assumed to be recoilless and follow eikonal
trajectories.  Therefore, their momenta do not change in the
interaction and the transverse component of the fermionic current is
zero.  The ``+'' and ``--''--components will still be $\delta$
functions on the light cone and in transverse direction, as was the
case at order $g$. The only effect of the collision on the valence
charges is a ``rotation'' of their color, as soon as a charge ``hits''
the field of the other charge at the collision point.  This
consideration leads us to the conclusion $J_+^{(3)} \sim \delta (x_- -
x_{-i})\, \theta (x_+ - y_{+j}) \, \delta({\underline x}-{\underline
x}_i)$ and $J_-^{(3)} \sim \theta (x_- - x_{-i}) \, \delta (x_+ -
y_{+j})\, \delta({\underline x}-{\underline y}_j)$. The correct
coefficients are found from Eq.\ (\ref{cons}):
\begin{mathletters}\label{current3}
\begin{eqnarray}
J_+^{(3)} & = & - {g^3 \over {2 \pi}} \, f^{abc}\, T^a \, (T_i^b)\,
(\tilde{T}_j^c) \, \delta (x_- - x_{-i})\, \theta (x_+ - y_{+j}) \,
\delta({\underline x}-{\underline x}_i) \, \ln (|{\underline x}_i -
{\underline y}_j| \lambda) \, \,\, , \\ 
J_-^{(3)} & = & {g^3 \over {2
\pi}} \, f^{abc}\, T^a \, (T_i^b)\, (\tilde{T}_j^c) \, \theta (x_- - x_{-i})
\, \delta (x_+ - y_{+j})\, \delta({\underline x}-{\underline y}_j)\, \ln
(|{\underline x}_i - {\underline y}_j| \lambda) \, \,\, , \\
{\underline J}^{(3)} & = & 0 \,\,\, .
\end{eqnarray}
\end{mathletters}
It can be shown that this current is consistent with the eikonal
scattering limit of either the lowest order QCD diagrams for gluon
radiation from two colliding color charges (cf.\ also Sect.\ IIIB below)
or Wong's equations \cite{wong} for the collision of two classical color 
charges.

For one charge, say $i$, ``hitting'' an ensemble of charges $j$, the
color of charge $i$ ``rotates'' each time it hits the field of one of
the charges $j$ in the ensemble. The resulting current is simply the
sum over $j$ of the terms on the right-hand side of (\ref{current3}).
Similarly, the collision of an ensemble with an ensemble simply adds
another summation over $i$. The actual nuclear collision in our
approach is only slightly more complex in that one has to account for
the presence of antiquarks as well.

  The equations of motion for the next-to-lowest order gluon field are
now obtained by inserting the lowest-order results together with the
current (\ref{current3}) into the right-hand side of Eq.\
(\ref{eom3}):
\begin{mathletters}\label{eom4}
\begin{eqnarray}
\Box\, A_+^{(3)a} & = & {g^3 \over { (2 \pi)^2 }} \, f^{abc}\, (T_i^b)
 \, (\tilde{T}_j^c) \left( - \,\, 2\pi\, \ln (|{\underline x}_i -
 {\underline y}_j| \lambda) \, \delta (x_- - x_{-i})\, \theta (x_+ -
 y_{+j}) \, \delta({\underline x}-{\underline x}_i) \right.  \\ & &
 \hspace*{3.2cm} \left. + \,\, \ln (|{\underline x} - {\underline
 x}_i| \lambda) \, \ln (|{\underline x} - {\underline y}_j| \lambda)
 \, \partial_+ \delta(x_- - x_{-i})\, \delta (x_+ - y_{+j})
 \right)\,\, , \nonumber \\
\Box\, A_-^{(3)a} & = & {g^3 \over { (2 \pi)^2 }} \, f^{abc}\, (T_i^b)
 \, (\tilde{T}_j^c) \left( \,\, 2\pi\, \ln (|{\underline x}_i -
 {\underline y}_j| \lambda) \, \theta (x_- - x_{-i})\, \delta (x_+ -
 y_{+j}) \, \delta({\underline x}-{\underline y}_j) \right.  \\ & &
 \hspace*{3.2cm}\left. - \,\, \ln (|{\underline x} - {\underline x}_i|
 \lambda) \, \ln (|{\underline x} - {\underline y}_j| \lambda) \,
 \delta(x_- - x_{-i})\, \partial_- \delta (x_+ - y_{+j}) \right) \,\,
 , \nonumber \\
\Box\, {\underline A}^{(3)a} & = & {g^3 \over { (2 \pi)^2 }}\,
 f^{abc}\, (T_i^b) \, (\tilde{T}_j^c) \, \delta(x_- - x_{-i}) \,
 \delta(x_+-y_{+j}) \\ & & \hspace*{3.2cm} \times \,\, \left(\, \ln(
 |{\underline x}- {\underline y}_j| \lambda) \, \frac{{\underline x}-
 {\underline x}_i}{|{\underline x}- {\underline x}_i|^2} - \ln(
 |{\underline x}- {\underline x}_i| \lambda) \, \frac{{\underline x}-
 {\underline y}_j}{|{\underline x}- {\underline y}_j|^2} \right) \,\,
 . \nonumber
\end{eqnarray}
\end{mathletters}

Equations (\ref{eom4}) are linear differential equations, which
justifies why we were able to focus on one single collision between
two valence quarks first and later obtain the complete solution by
summing over all possible collisions between color charges. (It was
explained above how the corresponding sum over $i$ and $j$ and over
quarks and antiquarks appears in the fermionic current; for the
commutator term, its presence is obvious.) The linearity of
(\ref{eom4}) also allows us to compute the solution simply by the
method of Green functions.  Since the classical solution obeys
causality, we have to use the retarded Green function:
\begin{equation}
A_{\mu}^{(3)} (x) = \int d^4 x'\, G_r (x - x')\, \tilde{J}_{\mu} (x'),
\label{clsol3}
\end{equation}
where $\tilde{J}_{\mu} \equiv J_\mu^{(3)} + ig\, [A^{(1)\nu} ,
\partial_\nu A_\mu^{(1)} + F_{\nu \mu}^{(1)} ] $ is given explicitly
by the right-hand side of Eq.\ (\ref{eom4}).  The retarded Green function reads
in coordinate and momentum space \cite{itzykson}:
\begin{equation}
G_r(x) = \frac{1}{2 \pi} \, \theta(t)\, \delta(x^2)\,\, , \,\,\,\,
\tilde{G}_r(k) = - \,\frac{1}{k^2 + i \epsilon k_0 }\,\, .
\label{Green}
\end{equation}
Formulae (\ref{eom4} -- \ref{Green}) provide the classical gluon field
to order $g^3$. Note that the above perturbative solution scheme 
renders the equations of motion linear at {\em each\/} successive order 
in $g$. In principle, one can therefore use the method of Green functions
to construct the classical solution to arbitrary order in $g$.
We shall argue below, however, that already at order $g^5$ quantum
effects become important and the classical approach breaks down.
Before we compute the solution to order $g^3$ explicitly, let us draw a
connection to the perturbative solution via Feynman diagrams.

\subsection{Connection to Feynman diagrams}

  Let us write the right-hand side of Eq.\ (\ref{clsol3}) in momentum
representation,
\begin{equation} \label{clsol4}
A_{\mu}^{(3)a} (x) = - \int \frac{ d^4 k}{(2 \pi)^4}\, \frac{e^{- i k
 \cdot x} }{k^2 + i \epsilon k_0 } \, \tilde{J}^a_{\mu}(k)\,\, ,
\end{equation}
where 
\begin{mathletters} 
\begin{eqnarray}
\tilde{J}_{+}^{a} (k) & = & \frac{g^3}{ (2 \pi)^2} \, f^{abc}\,
 (T_i^b) \, (\tilde{T}_j^c)\, e^{i (k_+ x_{-i} + k_- y_{+j} -
 \underline{k} \cdot \underline{y}_j)} \,\int d^2 \underline{q} \,
 e^{-i \underline{q} \cdot (\underline{x}_i - \underline{y}_j)} \,
 \frac{1}{(\underline{k} - \underline{q})^2} \, \left[ \frac{ i }{k_-
 + i \epsilon} - \frac{ i k_+ }{\underline{q}^2} \right] 
\,\, , \\
 \tilde{J}_{-}^{a} (k) & = & - \frac{g^3}{ (2 \pi)^2} \, f^{abc}\,
 (T_i^b) \, (\tilde{T}_j^c)\, e^{i (k_+ x_{-i} + k_- y_{+j} -
 \underline{k} \cdot \underline{y}_j)} \,\int d^2 \underline{q} \,
 e^{-i \underline{q} \cdot (\underline{x}_i - \underline{y}_j)} \,
 \frac{1}{\underline{q}^2} \, \left[ \frac{ i }{k_+ + i \epsilon} -
 \frac{i k_- }{ (\underline{k} - \underline{q})^2} \right] 
\,\, , \\
 \underline{\tilde{J}}^a (k) & = & \frac{g^3}{ (2 \pi)^2}
 \,f^{abc}\,(T_i^b) \, (\tilde{T}_j^c)\, e^{i (k_+ x_{-i} + k_- y_{+j}
 - \underline{k} \cdot \underline{y}_j)} \,\int d^2 \underline{q} \,
 e^{-i \underline{q} \cdot (\underline{x}_i - \underline{y}_j)} \,
 \frac{i (2\underline{q} - \underline{k})}{ \underline{q}^2 (
 \underline{k}- \underline{q})^2} \,\,\, .\label{tilJc}
\end{eqnarray} \label{tilJ}
\end{mathletters}
In order to derive this, we have made repeated use of 
\begin{equation} \label{ident}
\int { d^2 \underline{q} \over { (2 \pi)^2} }\,\, e^{ i \underline{q}
 \cdot {\underline{x}} } \,\, {1 \over {\underline{q}^2} } = - {1 \over {2
 \pi} } \ln (|\underline{x}| \lambda)
\end{equation}
and, for (\ref{tilJc}), of the transverse gradient of this equation.

\begin{figure}
\begin{center}
\epsfxsize=15cm
\epsfysize=3cm
\leavevmode
\hbox{ \epsffile{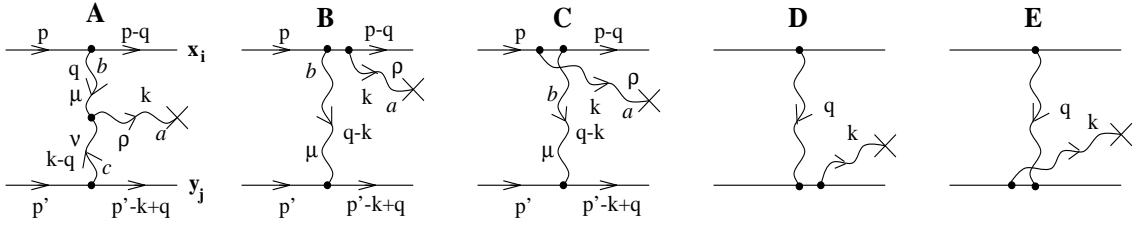}}
\end{center}
\caption{Diagrams contributing to the gluon field at order $g^3$.}
\label{diag}
\end{figure}

  The diagrams giving the classical gluon field \cite{yuri} at order $g^3$ are
shown in Fig.\ \ref{diag}. The cross at the end of the gluon
line denotes the space--time point $x$ where we measure the field.
The upper quark line corresponds to the first charge (fixed
coordinates $x_{-i},\, \underline{x}_i$), the lower one corresponds to
the second charge (at fixed $y_{+j}, \, \underline{y}_j$).  Therefore,
the momentum on the upper line has a large ``+''--component, the
momentum on the lower line has a large ``--''--component.  The gluon
field to order $g^3$ also includes graphs where the two quarks do not
interact. These diagrams are not shown in Fig.\ \ref{diag}, since they
are not part of the classical gluon field and do not contribute to
gluon production (they vanish once we take the emitted gluon line to
be on-shell).

We take the gluon-fermion vertices in the eikonal approximation.  In
the standard calculation of the diagrams, the emitted gluon line
corresponds to a gluon (Feynman) propagator $-i/(k^2 + i \epsilon)$
times a phase $e^{-i k \cdot x}$ (since we shall ultimately transform
the diagrams into coordinate space). 
This term is common to all diagrams in Fig.\ \ref{diag}.  
However, due to the regularization of the Feynman
propagator, all diagrams corresponding to gluon {\em absorption\/}
(instead of emission) are automatically included in this
calculation. That means, the usual Feynman diagrams yield an acausal
result. In order to establish correspondence to the classical result,
where there is only gluon emission, we replace the Feynman propagator
by the {\em retarded\/} propagator $-i/(k^2 +i \epsilon k_0)$ for the
gluon field measured at $x$.

Similarly, due to the Feynman regularization of the fermion
propagator, all diagrams $B - E$ contain contributions where the gluon
is emitted {\em prior\/} to the one--gluon exchange.  To ensure
causality, we have to use a {\em retarded\/} fermion propagator when
calculating the diagrams $B$ and $D$ and an {\em advanced\/} fermion
propagator in diagrams $C$ and $E$ which renders the emission of the
gluon causal.

 We have the freedom to change the regularization of
propagators. Different choices of regularization do not influence the
physics. The difference between the retarded (or advanced) propagators
and the usual Feynman propagator is just a $\delta$ function of the
square of the four-momentum of the internal line. Therefore, for
graphs $B - E$ this difference is proportional to those parts of the
diagrams where the fermion line is on-shell. But these parts do not
contribute to real gluon production, since once we put the outgoing
gluon on-shell, they vanish.  The regularization of the internal gluon
lines of the diagrams in Fig.\ \ref{diag} turns out to be of no
importance for the actual calculation.

After clarifying the regularization of the propagators, we compute the
diagrams according to the usual Feynman rules in covariant gauge. Let
us denote the $\rho$th component of the gluon field from a diagram $X$
by $X|_{\rho}$, where $X= A,B,C,D,$ or $E$.  Diagram $A$ is non-zero
for all values of $\rho$, while for $B$ and $C$ only the
``+''--component and for $D$ and $E$ only the ``--''--component are
non-vanishing.  After a lengthy, but straightforward calculation we
compare with Eqs.\ (\ref{clsol4}) and (\ref{tilJ}) to obtain the
identities
\begin{mathletters}
\begin{eqnarray}
A_+^{(3)a} (x) = (A+B+C) |_{\rho = +}\,\, , \\ 
A_-^{(3)a} (x) = (A+D+E) |_{\rho = -}\,\, , \\
\underline{A}^{(3)a} (x) = A |_{\rho = \perp}\,\, .
\end{eqnarray}
\end{mathletters}
We see that the calculation of the diagrams yields exactly the gluon
field (\ref{clsol4}) obtained from the classical solution of the
Yang--Mills equations.  That proves the correspondence of the
classical field to the diagrams in Fig.\ \ref{diag}. The diagram $A$
arises from the commutator term on the right-hand side of Eq.\
(\ref{eom3}), whereas graphs $B - E$ arise from the fermionic current.
   
\begin{figure}
\begin{center}
\epsfxsize=8cm
\epsfysize=2.5cm
\leavevmode
\hbox{ \epsffile{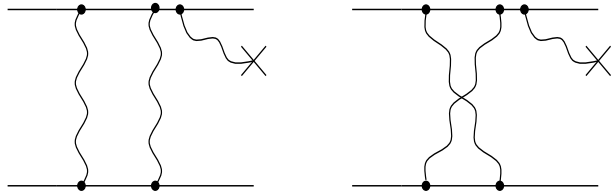}}
\end{center}
\caption{Graphs contributing to the gluon production at order $g^5$.}
\label{high}
\end{figure}

    The order $g^3$ is the limit of applicability of the classical
approach to the problem of gluon production. At order $g^5$ one can
construct diagrams which contribute to gluon production, but cannot be
obtained classically. An example of this type is shown in Fig.\
\ref{high}.

The two--gluon exchange contribution (Fig.\ \ref{high}) calculated
using the traditional Feynman regularization of the propagators cannot
be obtained from the classical equations of motion. The classical
approach would give the part of the diagram where the two exchanged
gluons are in a color singlet state. That corresponds to the internal
fermion lines between the exchanged gluons being on-shell. This is
what one would obtain by iterating the procedure for the
determination of the classical field outlined above to higher
orders in the coupling.  The color octet combination of the two
gluons, being the lowest order correction to the gluon's Regge
trajectory, is a pure ``quantum'' effect. 
A different regularization
of the propagators cannot make this contribution ``classical''. 
One can see that by a direct calculation of the
two--gluon exchange diagrams in the octet channel. The resulting
contribution is proportional to $\ln \, s$, where $s$ is the
center of mass energy of the colliding quarks [see \cite{Lipatov} and
references mentioned there]. That is, it depends on the longitudinal
momenta of the quarks. However, the classical calculation is just
an iteration of the one--gluon exchange diagram and thus not able to
provide such a logarithmic longitudinal dependence.
Therefore, the diagrams in Fig.\ \ref{high} are not classical.
(After all,
it would be quite surprising if the gluon's Regge trajectory turns out to
be a classical concept.)

\subsection{Classical gluon field in coordinate space}

   Let us now explicitly compute the gluon field (\ref{clsol3})
in coordinate space.
We decompose the ``+''--component of the field as
\begin{equation}
A_+^{(3)a} = a_1 +a_2\,\,\, ,
\end{equation}
where $a_1$ is straightforwardly computed:
\begin{eqnarray}
a_1 & = & - {g^3 \over { 2 \pi }} \, f^{abc}\, (T_i^b) \,
 (\tilde{T}_j^c)\int d^4 x' \, G_r (x-x')\, \ln (|{\underline x}_i -
 {\underline y}_j| \lambda) \, \delta (x'_- - x_{-i})\, \theta (x'_+ -
 y_{+j}) \, \delta({\underline x}'-{\underline x}_i) \nonumber \\ 
& =
 &- {g^3 \over {2 (2 \pi)^2 }}\, f^{abc}\, (T_i^b) \, (\tilde{T}_j^c)
 \, {1 \over {x_- - x_{-i}}}\, \theta (x_- - x_{-i})\, \theta \left(
 x_+ - y_{+j} - {|{\underline x} - {\underline x}_i|^2 \over {2(x_- -
 x_{-i})}} \right) \ln (|{\underline x}_i - {\underline y}_j|
 \lambda)\,\,\, .
\end{eqnarray}
The second term can be written in the form
\begin{eqnarray}
a_2 & = & {g^3 \over { (2 \pi)^2 }} \, f^{abc}\, (T_i^b) \,
 (\tilde{T}_j^c) \int d^4 x' \, G_r (x-x') \, \ln (|{\underline x}' -
 {\underline x}_i| \lambda) \, \ln (|{\underline x}' - {\underline
 y}_j| \lambda) \, \partial'_+ \delta(x'_- - x_{-i})\, \delta (x'_+ -
 y_{+j}) \nonumber \\
\label{a2}
 & =  & {g^3 \over {2 (2 \pi)^2 }} \, f^{abc}\, (T_i^b) \, (\tilde{T}_j^c)\,
 \partial_+ [\theta (x_- - x_{-i}) \, \theta ( x_+ - y_{+j})\, {\cal J} ]\,\, ,
\end{eqnarray}
with
\begin{equation} \label{intJ}
{\cal J} \equiv \int {{d^2 {\underline q}\, d^2 {\underline l} \over
{(2 \pi)^2}}\, e^{- i {\underline q} \cdot ({\underline x} -
{\underline y}_j ) - i {\underline l} \cdot ({\underline x} -
{\underline x}_i ) } \, \frac{1}{{\underline l}^2 \, {\underline
q}^2}} \, J_0 (|{\underline q} + {\underline l}| \tau)\,\,\, ,
\end{equation}
where $\tau = \sqrt{ 2 (x_- - x_{-i})( x_+ - y_{+j}) }$. The explicit
evaluation of the integral $\cal J$ is referred to Appendix A. The
final result is:
\begin{eqnarray}
{\cal J} = \ln (\xi_> \lambda)\, \ln (\eta_> \lambda) + \frac{1}{4}
\left[ \mbox {Li}_2 \left( e^{i \alpha} \frac{\xi_< \eta_<}{\xi_>
\eta_>} \right) + \mbox {Li}_2 \left( e^{- i \alpha} \frac{\xi_<
\eta_<}{\xi_> \eta_>} \right) \right]\,\, ,
\end{eqnarray}
where $\xi_{> (<)} = \mbox {max} (\mbox {min}) (|{\underline x} -
{\underline x}_i|, \tau) \, , \eta_{> (<)} = \mbox {max} (\mbox {min})
(|{\underline x} - {\underline y}_j|, \tau)$, and $\alpha$ is the
angle between ${\underline x} - {\underline x}_i $ and ${\underline x}
- {\underline y}_j$. $\lambda$ plays the same role as in Eq.\
(\ref{clsol1}).  $\mbox {Li}_2 (z)$ is the dilogarithm (also known as
Spence's function). With the final expression for $a_2$ we obtain:
\begin{eqnarray}\label{clsor31}
\lefteqn{A_+^{(3)a}(x) = - {g^3 \over {2 (2 \pi)^2 }} \, f^{abc}\, (T_i^b) \,
(\tilde{T}_j^c) \, \left[ {1 \over {x_- - x_{-i}}} \, \theta (x_- -
x_{-i})\, \theta \left( x_+ - y_{+j} - {|{\underline x} - {\underline
x}_i|^2 \over {2(x_- - x_{-i})}} \right) \ln (|{\underline x}_i -
{\underline y}_j| \lambda) \right. } \nonumber \\ & - & \delta (x_- -
x_{-i})\, \theta ( x_+ - y_{+j}) \, \ln (|{\underline x} - {\underline
x}_i| \lambda) \, \ln (|{\underline x} - {\underline y}_j|\lambda) -
\theta (x_- - x_{-i}) \, \theta ( x_+ - y_{+j})\, \partial_+ [\ln
(\xi_> \lambda) \, \ln (\eta_> \lambda)] \nonumber \\ & + &
\left. \frac{1}{4}\, \theta (x_- - x_{-i}) \, \theta ( x_+ - y_{+j})\,
(\partial_+ \ln r ) \, \ln (1 - 2 r \cos \alpha + r^2) \right]\,\,\,,
\end{eqnarray}
where $r = (\xi_< \eta_<)/(\xi_> \eta_>)$.

   The integrations for $A_-^{(3)a}$ are done similarly with the
result:
\begin{eqnarray}\label{clsor32}
\lefteqn{A_-^{(3)a}(x) = {g^3 \over {2 (2 \pi)^2 }} \, f^{abc}\, (T_i^b) \,
(\tilde{T}_j^c)\, \left[ {1 \over {x_+ - y_{+j}}}\, \theta (x_+ -
y_{+j})\, \theta \left( x_- - x_{-i} - {|{\underline x} - {\underline
y}_j|^2 \over {2(x_+ - y_{+j})}} \right) \ln (|{\underline x}_i -
{\underline y}_j| \lambda) \right. } \nonumber \\ & - & \delta (x_+ -
y_{+j})\, \theta ( x_- - x_{-i})\, \ln (|{\underline x} - {\underline
x}_i| \lambda)\, \ln (|{\underline x} - {\underline y}_j| \lambda) -
\theta (x_- - x_{-i})\, \theta ( x_+ - y_{+j})\, \partial_- [\ln
(\xi_> \lambda)\, \ln (\eta_> \lambda)] \nonumber \\ & + &
\left. \frac{1}{4}\, \theta (x_- - x_{-i})\, \theta ( x_+ - y_{+j})\,
(\partial_- \ln r ) \,\ln (1 - 2 r \cos \alpha + r^2) \right]\,\,\, .
\end{eqnarray}
For ${\underline A}^{(3)a}$ we find
\begin{eqnarray}\label{clsor33}
{\underline A}^{(3)a} (x) & = & - {g^3 \over {2 (2 \pi)^2 }} \, f^{abc}\,
(T_i^b) \, (\tilde{T}_j^c)\, \theta (x_- - x_{-i}) \, \theta ( x_+ -
y_{+j}) \, [(\nabla_i - \nabla_j)\, {\cal J} ]\nonumber \\ 
& = & -
{g^3 \over {2 (2 \pi)^2 }} \, f^{abc}\, (T_i^b) \, (\tilde{T}_j^c)\,
\theta (x_- - x_{-i})\, \theta ( x_+ - y_{+j}) \, \left[ \frac{}{} (\nabla_i -
\nabla_j)\, [\ln (\xi_> \lambda) \ln (\eta_> \lambda)]  
\right. \nonumber \\ 
&  & \left. - \frac{1}{4}
[(\nabla_i - \nabla_j) \ln r]\, \ln (1 - 2 r \cos \alpha + r^2) 
- \frac{i}{4} \, [(\nabla_i -
\nabla_j) \alpha ]\, \ln \left( \frac{1 - r e^{i \alpha}}{1 - r e^{- i
\alpha}} \right) \right]\,\,\, .
\end{eqnarray}
As expected, the solution to order $g^3$ is causal, i.e., it is
non-vanishing {\em only\/} in the forward light-cone.

To obtain the complete solution for the nucleus--nucleus collision we
simply sum over all charges in the nuclei, i.e., as discussed above,
we sum over all nucleons $i$ and $j$ and over quarks and antiquarks
inside the nucleons. If we denote the solution of the scattering
problem of two single charges with coordinates $x_i$ and $y_j$ found
above by $A^{(3)a}_\mu (x,x_i,y_j)$ we obtain
\begin{eqnarray} \label{finsol}
A_{\mu}^{(3)a} (x) = \sum_{i,j = 1}^{A_1 , A_2} [A_{\mu}^{(3)a} (x,x_i,y_j) -
A_{\mu}^{(3)a} (x,x'_i,y_j) - A_{\mu}^{(3)a} (x,x_i,y'_j) + A_{\mu}^{(3)a}
(x,x'_i,y'_j) ]\,\,\, .
\end{eqnarray}
The antiquark coordinates are marked with a prime.  The relative signs
emerge from the fact that antiquarks have the same color charge as
quarks, but with opposite sign.

\section{The radiated field energy, number, and multiplicity distributions}

  In order to determine the radiated field energy we start from Eq.\
(\ref{clsol4}), and, for simplicity, discuss the case of a single
collision first. Note that a part of the solution corresponds just to
a change of the field carried by the charge due to the collision (the
color ``rotation''), and not to the radiated gluon field. That part is
most easily isolated by a contour integration in the complex
$k_0$--plane in Eq.\ (\ref{clsol4}), it arises from the pole $k_\pm =
-i \epsilon$ in Eq.\ (\ref{tilJ}):
\begin{mathletters}
\begin{eqnarray}
A_{+ {\rm \,charge}}^{(3)a}(x) & = & \frac{g^3}{ (2\pi)^2}\, f^{abc}\,
(T_i^b)\, (\tilde{T}_j^c) \, \ln (|{\underline x}_i - {\underline
y}_j| \lambda) \, \ln (|{\underline x} - {\underline x}_i| \lambda) \,
\delta(x_- - x_{-i})\, \theta (x_+ - y_{+j}) \, , \\ 
A_{- {\rm \,
charge}}^{(3)a} (x) & = & - \frac{g^3}{ (2\pi)^2}\, f^{abc}\,
(T_i^b)\, (\tilde{T}_j^c) \, \ln (|{\underline x}_i - {\underline
y}_j| \lambda) \, \ln (|{\underline x} - {\underline y}_j| \lambda) \,
\theta(x_- - x_{-i}) \, \delta(x_+ - y_{+j}) \, , \\ 
{\underline A}_{\rm \, charge}^{(3)a} (x) & = & 0 \,\,\, .
\end{eqnarray}
\end{mathletters}
The radiated gluon field arises from the poles of the retarded
propagator in Eq.\ (\ref{clsol4}), i.e., it corresponds to on-shell
gluons, as one would expect:
\begin{equation}
A_{\mu{\rm \,rad}}^{(3)a} (x) = \theta(t - t_{ij})\, \int d \tilde{k}
\left[ \, i\, \tilde{J}_{\mu}^a(\omega,{\bf k}) \,\, e^{-i \omega t +
i{\bf k} \cdot {\bf x}} + {\rm c.c.} \right] \,\, ,
\end{equation}
where $t_{ij} \equiv (x_{-i} + y_{+j})/\sqrt{2}$ is the time when the
collision happens and $ d\tilde{k} \equiv d^3 {\bf k}/[(2 \pi)^3 2
\omega]$, with $\omega = |{\bf k}| $ \cite{itzykson}. Obviously, this
field vanishes prior to the collision.

The (stationary part of the) radiated field energy is \cite{itzykson}
\begin{equation}
{\cal H} = \int {\rm d} \tilde{k} \, \omega \left[ \,
\underline{\tilde{J}}^a (\omega, {\bf k}) \cdot \underline{\tilde{J}}^{a*} 
(\omega, {\bf k}) - 
\tilde{J}^a_+ (\omega, {\bf k}) \, \tilde{J}^{a*}_- (\omega, {\bf k})
- \tilde{J}^a_- (\omega, {\bf k})\, \tilde{J}^{a*}_+(\omega, {\bf k}) 
\right] \,\, .
\end{equation}
In the case of a nuclear collision $\tilde{J}_{\mu}^a(k)$, Eq.\
(\ref{tilJ}), becomes more complicated: there is an additional sum
over nucleons $i$ and $j$ and over quarks and antiquarks inside these
nucleons.  Inserting the resulting expression, for times {\em after\/}
the {\em last\/} parton--parton collision we arrive at
\begin{eqnarray}\label{diagsq}
{\cal H} & = & 4\, \frac{g^6}{(2\pi)^4} \,
f^{abc}\, f^{ade} \int {\rm d} \tilde{k}\, \omega \,\, 
\frac{1}{{\underline k}^2}   
\int d^2 {\underline q}_1 \, d^2 {\underline q}_2 \,
\frac{{\underline q}_1 \cdot {\underline q}_2 \, {\underline k}^2 + 
{\underline q}_1^2\, {\underline q}_2^2 - {\underline q}_1^2\, 
{\underline k} \cdot {\underline q}_2 -{\underline q}_2^2 \, 
{\underline k} \cdot {\underline q}_1}{ {\underline q}_1^2\,\, 
{\underline q}_2^2 \,\, (\underline{k} - \underline{q}_1)^2\,\, 
(\underline{k}-\underline{q}_2)^2} \,\, \\
& \times & \sum_{i,k=1}^{A_1} \sum_{j,l=1}^{A_2} 
(T_i^b) \, (\tilde{T}_j^c) \, (T_k^d) \, (\tilde{T}_l^e) \,\, 
 {\cal P}(k_+,\underline{q}_1;x_i) \, 
 {\cal P}(k_-,\underline{k} - \underline{q}_1;y_j)\,
 {\cal P}^*(k_+, \underline{q}_2;x_k)\, 
 {\cal P}^*(k_-, \underline{k}- \underline{q}_2,y_l) \,\, ,
 \nonumber
\end{eqnarray}
where now $k_\pm \equiv (\omega \pm k^z)/\sqrt{2}$ and
\begin{equation}
{\cal P}(k_+,\underline{q};x_i) \,\, \equiv \,\, e^{ i k_+ x_{-i} - i
{\underline q} \cdot {\underline x}_i }\,\, - \,\, e^{ i k_+ x_{-i}' -
i {\underline q} \cdot {\underline x}_i'} \,\, .
\end{equation}

In order to achieve color neutrality in the initial state, we have to
average over all possible color orientations in the color space of the
individual nucleons. With ${\rm tr}\,[\, (T_i^b) \,(\tilde{T}_j^c)\,
(T_k^d)\, (\tilde{T}_l^e)\,] = \delta_{ik} \, \delta^{bd} \delta_{jl}
\,\delta^{ce} /4$, and $f^{abc} \, f^{abc} = N_c (N_c^2-1)$ this
yields
\begin{equation} \label{H}
\frac{1}{N_c^2}\,\, {\rm tr} [ {\cal H}] =  
\frac{g^6}{(2\pi)^4}\, \frac{N_c^2-1}{N_c} \,
\int d \tilde{k} \, \frac{\omega}{{\underline k}^2} \int
d^2 {\underline q}_1 \, d^2 {\underline q}_2 \, 
\frac{{\underline q}_1 \cdot {\underline q}_2 \, {\underline k}^2
 + {\underline q}_1^2\, {\underline q}_2^2 - {\underline q}_1^2\, 
{\underline k} \cdot {\underline q}_2
-{\underline q}_2^2 \, {\underline k} \cdot {\underline q}_1}{ 
{\underline q}_1^2\,\,  {\underline q}_2^2 
\,\, ({\underline k - q}_1)^2\,\, ({\underline k-q}_2)^2}\,\, 
\tilde{\cal P} (k,\underline{q}_1, \underline{q}_2) \, ,
\end{equation}
where 
\begin{equation}
\tilde{\cal P} (k,\underline{q}_1, \underline{q}_2) \equiv
\sum_{i=1}^{A_1} {\cal P} (k_+, \underline{q}_1;x_i)\, {\cal P}^*(k_+,
\underline{q}_2;x_i) \sum_{j=1}^{A_2} {\cal P} (k_-, \underline{k}-
\underline{q}_1;y_j) \, {\cal P}^*(k_-, \underline{k}-
\underline{q}_2;y_j) \,\, .
\end{equation}
For further evaluation we introduce the center-of-mass coordinates of
nucleon $i$, $(X_{-i}, {\underline X}_i)$, and nucleon $j$, $(Y_{+j},
{\underline Y}_j)$, and relative coordinates $\Delta x_{-i}, \, \Delta
{\underline x_i}, \, \Delta y_{+j},\, \Delta {\underline y}_j$, such
that
\begin{mathletters}
\begin{eqnarray}
x_{-i} = X_{-i} + \frac{\Delta x_{-i}}{2}\,\, , \,\,\, x_{-i}' =
X_{-i} - \frac{\Delta x_{-i}}{2} \,\,\, & , & \,\,\, y_{+j} = Y_{+j} +
\frac{\Delta y_{+j}}{2} \,\, , \,\,\, y_{+j}' = Y_{+j} - \frac{\Delta
y_{+j}}{2}\,\, , \\ {\underline x}_i = {\underline X}_i + \frac{\Delta
{\underline x}_i}{2} \,\, , \,\,\,{\underline x}_i' = {\underline X}_i
- \frac{\Delta {\underline x}_i}{2} \,\, & , & \,\,\, {\underline y}_j
= {\underline Y}_j + \frac{\Delta {\underline y}_j}{2} \,\,, \,\,\,
{\underline y}_j' = {\underline Y}_j - \frac{\Delta {\underline
y}_j}{2} \,\, .
\end{eqnarray}
\end{mathletters}
Here we assumed that the positions of quark and antiquark in a nucleon are
symmetric with respect to the center of the nucleon. This assumption
is not crucial. The calculations can also be done for the general case
of arbitrary positions of quarks and antiquarks in the nucleons. For
the ``symmetric'' case the phase factor becomes
\begin{eqnarray}
\tilde{\cal P}(k,\underline{q}_1, \underline{q}_2) & = &
\sum_{i=1}^{A_1} e^{-i ({\underline q}_1 - {\underline q}_2) \cdot
{\underline X}_i } \left[ \,\, e^{-i ({\underline q}_1 - {\underline
q}_2) \cdot \Delta {\underline x}_i/2} - e^{ik_+ \Delta x_{-i}
-i({\underline q}_1 + {\underline q}_2) \cdot \Delta {\underline
x}_i/2} + {\rm c.c.}\,\, \right] \nonumber \\ & \times &
\sum_{j=1}^{A_2} e^{i ({\underline q}_1 - {\underline q}_2) \cdot
{\underline Y}_j } \left[ \,\, e^{i ({\underline q}_1 - {\underline
q}_2) \cdot \Delta {\underline y}_j/2} - e^{ik_- \Delta y_{+j} -i
{\underline k} \cdot \Delta {\underline y}_j + i({\underline q}_1 +
{\underline q}_2) \cdot \Delta {\underline y}_j/2} + {\rm c.c.}\,\,
\right] \,\, .
\end{eqnarray}
We shall now first average $\tilde{\cal P}$ over the longitudinal
positions of quarks and antiquarks inside the nucleons. The nuclei
are highly Lorentz-contracted in the longitudinal direction. To
simplify the subsequent discussion,
we take them and the nucleons inside to be
cylindrical. Each nucleon is then a cylinder of radius $a$ and
length $2a/\gamma$, oriented along the $z$--axis, where
$\gamma$ is the Lorentz-factor in the CM frame of the collision.
(The longitudinal extension of the nucleons is not important, it
will drop out in the following anyway.) With
\begin{equation} \label{longave}
\frac{\gamma}{\sqrt{2} a} \int_{-a/\sqrt{2}\gamma}^{a/\sqrt{2}\gamma}
d\left( \frac{ \Delta x_{-i}}{2}\right) \, e^{ik_+ \Delta x_{-i}} =
\frac{\sin[ k_+ \sqrt{2} a/ \gamma]}{ k_+ \sqrt{2} a/ \gamma}
\rightarrow 1 \,\,\,\,\,\,\, (\gamma \rightarrow \infty) \,\, ,
\end{equation}
and an analogous relation for the average over $\Delta y_{+j}$, the
longitudinal momentum dependence vanishes from the phase factor. 

Let us now average over the transverse positions ${\underline X}_i,
{\underline Y}_j$ of the nucleons inside a nuclear transverse
area. For a cylindrical nucleus, the transverse area which we average
over is independent of the longitudinal position of the individual
nucleon and equal to $\pi R_{1 (2)}^2$, $R_{1 (2)}$ being the
transverse radius of nucleus 1 (2). Let us take nucleus 1 to be the
larger of the two nuclei. Since $R_1$ is by far larger than the
(inverse) momentum scales we are interested in, we may take to good
approximation $R_1 \rightarrow \infty$, and obtain
\begin{equation} \label{R}
\frac{1}{\pi R_1^2} \int d^2 {\underline X}_i \,\, e^{-i ({\underline
q}_1 - {\underline q}_2) \cdot {\underline X}_i} \simeq \frac{4
\pi}{R_1^2}\, \delta ({\underline q}_1 - {\underline q}_2)\,\, .
\end{equation}
With this result, the ${\underline Y}_j$--average (over the transverse
positions of nucleons in nucleus 2) is trivial.
The assumption of infinite transverse nuclear size greatly simplifies
the subsequent discussion. However, we are then no longer able to study
collisions at finite impact parameter. 

Finally, we average over the transverse dimension of the nucleons.
That integral is formally the same as (\ref{R}), only that $R_1$ is
replaced by $a$ and the momentum dependences of the various terms in
$\tilde{\cal P}$ are different. Now, however, $1/a$ is not small on
the momentum scale of interest. In fact, the ${\underline
q}_{1,2}$--integrations in (\ref{H}) are logarithmically divergent,
such that the treatment of the small momentum region is of some
importance. As we shall see, treating the average over the transverse
dimensions of the nucleons correctly leads to an infrared
regularization of these divergences via nucleonic form factors. 
The physical interpretation is
that on large (spatial) scales individual nucleons appear colorless
and do not emit gluons.

The average over the transverse dimension of a nucleon involves
typically integrals of the type \cite{GR}
\begin{equation}
\frac{1}{\pi a^2} \int d^2 \left( \frac{\Delta {\underline x}_i}{2}
\right)\,\, e^{i {\underline q}\cdot \Delta {\underline x}_i/2} =
\frac{ 2 \, J_1 (|{\underline q}|a)}{|{\underline q}| a}\,\, .
\end{equation}
For the average phase factor we thus obtain
\begin{equation}
\langle \, \tilde{\cal P} (k,\underline{q}_1, \underline{q}_2)\, \rangle
= 4\, A_1 A_2 \, \frac{4\pi}{R_1^2}\, \delta ({\underline q}_1 -
{\underline q}_2)\,\, \left(\, 1 - \frac{ 2\, J_1(2 |{\underline
q}_1|a)}{2|{\underline q}_1|a} \right)\,\, \left(\, 1 - \frac{ 2\,
J_1(2 |{\underline k}- \underline{q}_1|a)}{2|{\underline
k}-\underline{q}_1|a} \right) \,\,.
\end{equation}
Our final result for the invariant distribution of the
average energy is
\begin{equation} \label{Edist}
\frac{d \langle {\cal H} \rangle}{ dy\, d^2 {\underline k}} =
\pi\,\frac{g^6}{(2\pi)^6}\, \frac{N_c^2 -1}{N_c}\, \frac{4 A_1
A_2}{\pi R_1^2} \,\, \frac{\omega}{{\underline k}^2}\,\, \int d^2
{\underline q} \,\, \frac{F(|{\underline q}|a)}{{\underline q}^2} \,\,
\frac{F(|\underline{k}-\underline{q}|a)}{(\underline{k}-
\underline{q})^2} \,\, ,
\end{equation}
with the (longitudinal) gluon rapidity $y \equiv \ln [k_+/k_-]/2$ and
the ``form factor'' $F(x) = 1- 2\, J_1(2x)/2x$. For small
${\underline q}$ or ${\underline k}-\underline{q}$, $F$ regularizes
the infrared divergence of the integral. If we do not assume that the
quark and antiquark are symmetric with respect to the nucleon's
center and allow them to be anywhere in the nucleon, the ``form
factor'' becomes $F(x) = 1 - [2 \, J_1(x)/x]^2$.

We now compare our result to that of \cite{Alex}. In that work, the
{\em number distribution\/} of radiated gluons was related to the
average energy distribution via
\begin{equation}
\frac{ d{\cal N}}{ d y\, d^2 {\underline k}} \equiv \frac{1}{\omega} \,
\frac{ d \langle {\cal H}\rangle}{ dy\, d^2 {\underline k}} \,\, .
\end{equation}
As in \cite{Alex}, we consider a central collision of equal nuclei, $A_1 = A_2
=A$.  According to \cite{yuri}, the average transverse color charge
density is given by $\mu^2 = C_F A/(N_c \pi R^2)$ (for our case of a
``cylindrical'' nucleus), where $C_F \equiv (N_c^2 - 1)/(2\,
N_c)$. Then, with the transverse area $S_\perp \equiv \pi R^2$,
\begin{equation} \label{Ndist}
\frac{d {\cal N}}{ dy\, d^2 {\underline k}} = S_\perp \,
\frac{\pi}{C_F^2} \,\, \frac{2\,g^6 \mu^4}{(2\pi)^3 \pi}\, N_c
(N_c^2-1) \,\, \frac{1}{{\underline k}^2}\,\, \int \frac{d^2
{\underline q}}{(2 \pi)^2} \,\, \frac{F( |{\underline
q}|a)}{{\underline q}^2} \,\, \frac{F(
|\underline{k}-\underline{q}|a)}{(\underline{k}- \underline{q})^2}
\,\, .
\end{equation}
This has (apart from the form factors) the same momentum dependence as
the result of \cite{Alex} [Eq.\ (49)]. The prefactor, though different
from \cite{Alex}, agrees with recent results obtained by Gyulassy and McLerran
\cite{Miklos}. Note that due to the form factors, our result behaves
like $1/\underline{k}^2$ for small $|\underline{k}| \sim 1/a$. The
residual infrared divergence stems from our approximation of an
infinitely large nucleus in Eq.\ (\ref{R}). This divergence is
actually cut off by the finite (inverse) size of the nucleus
(i.e., by the nuclear form factor).  In this
range of momenta, however, our treatment ceases to be valid anyway,
since then $|\underline{k}| \sim 1/R \ll 1/a \sim \Lambda_{\rm QCD}$,
while one of our initial assumptions was $|\underline{k}| \gg
\Lambda_{\rm QCD}$.

The gluon {\em number\/} distribution (\ref{Ndist}) is {\em not\/}
directly accessible in the experiment. Experimentally, one measures
the (invariant) differential cross section for detecting a radiated
gluon, $d\sigma_{\rm rad}/dy\, d^2 {\underline k}$, divided by the
total cross section of the collision.  Equivalently, one can
divide the gluon {\em number\/} distribution (\ref{Ndist}) by the
total number of scattering events with one--gluon exchange
between the color charges in a (central) $A+A$--collision. This number
can be obtained quite analogously to the above derivation of the
number distribution for radiated gluons. Instead of Eq.\ (\ref{diagsq})
we start with the square of the one--gluon exchange amplitude. This
amplitude for two charges (whose coordinates are specified as in
Sect.\ III) is
\begin{equation}
-\, \frac{i\, g^2}{(2 \pi)^2} \, (T^a_i) \, (\tilde{T}^a_j) \, \int \,
\frac{d^2 {\underline q}}{{\underline q}^2 } \, \, e^{i{\underline q}
\cdot ({\underline y}_j - {\underline x}_i) }.
\end{equation}
Similar to the above, after averaging the modulus squared of this 
amplitude over the
positions of the quarks and antiquarks inside the nucleons and nucleons 
in the nucleus,
as well as over quark colors in the initial state, we obtain
\begin{equation}\label{nelsc}
{\cal N}_{\rm tot} = 4\, \alpha_S^2\, \frac{C_F}{2\, N_c}\, \frac{4\, A^2}{\pi
R^2} \int d^2 \underline{q}\, \left(\frac{F(|\underline{q}|a)}{
\underline{q}^2} \right)^2 \,\,\, .
\end{equation}
The final answer for the gluon {\em multiplicity\/} distribution is
\begin{equation}\label{final}
\frac{d\,n}{dy\, d^2 \underline{k}} = \frac{1}{{\cal N}_{\rm tot}}\, \frac{
d{\cal N}}{dy\, d^2 \underline{k}} = \frac{ N_c \, \alpha_S}{\pi^2 \,
{\underline k}^2} \, \int d^2 {\underline q} \,\, \frac{F(
|{\underline q}|a)}{{\underline q}^2} \,\, \frac{F(
|\underline{k}-\underline{q}|a)}{(\underline{k}- \underline{q})^2}
\,\, \left[ \int d^2 \underline{q}\, \left(\frac{F(|\underline{q}|a)}{
\underline{q}^2} \right)^2 \right]^{-1}\,\,.
\end{equation}
Expanding the form factor (in the ``symmetric'' case)
for small $a$,  $F(x) \simeq x^2/(2 + x^2/3)$ for $x \rightarrow 0$, 
and choosing $m_\rho^2 \equiv 24/a^2$,
this result coincides with that of \cite{gunion} [Eq.\ (21)], since
$C_A \equiv N_c = 3$.

  For the case of a nucleon in which the positions of quark and
antiquark are symmetric with respect to its center one can explicitly
perform the integration in Eq.\ (\ref{final}). The result reads:
\begin{equation} \label{final2}
\frac{d\,n}{dy\, d^2 \underline{k}} = \frac{ N_c \, \alpha_S}{\pi^2 \,
{\underline k}^2} \, \frac{12}{5} \left( - \frac{1+(\underline{k} a)^2}{
({\underline k}a)^4} \, \, \left( 1 - J_0 (2|{\underline k}|a)\right) +
\frac{2 + ({\underline k}a)^2}{({\underline k}a)^4} \, \, 
J_2 (2|{\underline k}|a)\, + \,\, {_2}F_3 (1,1;2,2,2; - \underline{k}^2 a^2) 
\right).
\end{equation}
The details of the integration are given in Appendix B. For
$\underline{k} \rightarrow 0$, the expression in large parentheses
becomes $5/12$, as expected from (\ref{final}).

\section{Conclusions}

In this paper we have solved the classical Yang--Mills equations for a
collision of ultrarelativistic nuclei. Our solution assumes the nuclei
to have the form used in \cite{yuri} and the color charges to follow
eikonal trajectories during the collision. The solution is constructed
perturbatively in covariant (Lorentz) gauge to lowest order (Abelian
limit, order $g$) and next-to-lowest order ($g^3$). Via Feynman
diagrams we have clarified the connection between the classical and
quantum solution to order $g^3$. 

  We have established a limit for the classical description of gluon
production. From the discussion of the diagrams in Sect.\ IIIB follows
that the limit for the field is one gluon per nucleon, unless the
second gluon is the outgoing gluon. That means that in the
diagrammatic formulation of the problem there should be no more than
one gluon leaving each nucleon. In nuclear collisions, different from
the case of a single nucleus \cite{yuri'}, we cannot require the final
states of the nucleons to be color singlets. Therefore, here we cannot
apply color averaging which cancels quantum corrections [as described
in \cite{yuri'}] and puts internal fermion lines of graphs like in
Fig.\ \ref{high} on shell, allowing for a classical description. That way the
limit of the classical approach is reduced to one gluon per nucleon.

We have given explicit expressions for the gluon field in coordinate
space [Eqs.\ (\ref{clsor31} -- \ref{clsor33})] and calculated the
energy, number, and multiplicity distributions of radiated gluons to
order $g^3$ [Eqs.\ (\ref{Edist}, \ref{Ndist}, \ref{final})]. Our result
for the number distribution agrees with
\cite{Miklos} and the form factors of the individual nucleons can be cast
into a form such that the multiplicity distribution 
agrees with the result of \cite{gunion}.

The resulting gluon number and multiplicity distributions are
boost-invariant, a property which was a priori assumed in
\cite{Alex}. Boost--invariance is also assumed in popular hydrodynamic
models \cite{Bjorken} to describe the evolution of ultrarelativistic
nuclear collisions after (local) thermalization is established. To
answer the question whether thermalization actually happens and what
the respective time scales are, it is planned to utilize the solution
found here to study screening and damping in the radiation-produced
gluonic medium \cite{future}.

\section*{Acknowledgments}

The authors would like to thank S.\ Brodsky, M.\
Gyulassy, A.H.\ Mueller, B.\ M\"uller, S.\ Pratt, R.\ Venugopalan,
and J.\ Verbaarschot for
interesting discussions and helpful advice. The research of Yu.V.K.\
was sponsored in part by the U.S.\ Department of Energy under grant
DE-FG02-94ER-40819.  The work of D.H.R.\ was supported in part by the U.S.\
Department of Energy under Contract No.\ DE-FG02-96ER-40945.
\appendix

\section{}

  In this appendix we explicitly calculate the integral
\begin{eqnarray}
{\cal J} = \int {d^2 {\underline q} \, d^2 {\underline l} \over {(2
\pi)^2}}\, {e^{- i {\underline q} \cdot ({\underline x} - {\underline
y}_j ) - i {\underline l} \cdot ({\underline x} - {\underline x}_i )}
\over {{\underline l}^2 {\underline q}^2}} \, J_0 (|{\underline q} +
{\underline l}| \tau).
\end{eqnarray}
Using the addition theorem for Bessel functions ($\phi$ is the
angle between ${\underline q}$ and ${\underline l}$)
\begin{eqnarray}
J_0 (|{\underline q} + {\underline l}| \tau) = \sum_{k = - \infty}^{
\infty} (-1)^k e^{i k \phi} J_k (|{\underline q}| \tau) J_k
(|{\underline l}| \tau)
\end{eqnarray}
we obtain:
\begin{eqnarray}
{\cal J} = \sum_{k = - \infty}^{ \infty} (-1)^k \int {d^2
{\underline q} \, d^2 {\underline l} \over {(2 \pi)^2}} \, {e^{- i
{\underline q} \cdot ({\underline x} - {\underline y}_j ) - i
{\underline l} \cdot ({\underline x} - {\underline x}_i )} \over
{{\underline l}^2 {\underline q}^2}}\, e^{i k \phi} J_k (|{\underline
q}| \tau) \, J_k (|{\underline l}| \tau).
\end{eqnarray}
If $\phi_1$ is the angle between ${\underline l}$ and 
${\underline x} - {\underline x}_i$ , and $\phi_2$ is the angle 
between ${\underline q}$ and ${\underline x} - {\underline y}_j$,
we can express $\phi$ as 
\begin{equation}
\phi = \phi_2 - \phi_1 + \alpha,
\end{equation}
with $\alpha$ being the angle between ${\underline x} - {\underline
y}_j$ and ${\underline x} - {\underline x}_i$. All angles are taken
clockwise. Integrating over $\phi_1$ and $\phi_2$ we
obtain:
\begin{eqnarray}\label{bigj}
{\cal J} = \sum_{k = - \infty}^{\infty} e^{i k \alpha} \left(
\int_0^\infty \frac{dq}{q} J_k (q |{\underline x} - {\underline y}_j|)
J_k (q \tau) \right) \left( \int_0^\infty \frac{dl}{l} J_k (l
|{\underline x} - {\underline x}_i|) J_k (l \tau) \right).
\end{eqnarray}
For non-zero $k$ we can use the formula [see \cite{ww}]
\begin{eqnarray}
\int_0^\infty \frac{dl}{l}\, J_k (l |{\underline x} - {\underline x}_i|)\,
J_k (l \tau) = \frac{1}{2 k} \left( \frac{\xi_<}{\xi_>}\right)^k\,\, , \,\,
\,k > 0\,\, ,
\end{eqnarray}
where $\xi_{> (<)} = \mbox {max} (\mbox {min}) (|{\underline x} -
{\underline x}_i|, \tau) $, to obtain the series:
\begin{eqnarray}
\sum_{k = 1}^{\infty} ( e^{i k \alpha} + e^{- i k \alpha}
) \frac{1}{4 k^2} \left( \frac{\xi_< \eta_<}{\xi_> \eta_>}\right)^k,
\end{eqnarray}
with $\eta_{> (<)} = \mbox {max} (\mbox {min}) (|{\underline x} -
{\underline y}_j|, \tau)$. Using the definition of the dilogarithm,
\begin{equation}
\mbox {Li}_2 (z) \equiv \sum_{k = 1}^{\infty} {z^k \over {k^2}} \,\, ,
\end{equation}
we can rewrite this series as 
\begin{equation}
\frac{1}{4} \left[ \mbox {Li}_2 \left( e^{i \alpha} \frac{\xi_<
\eta_<}{\xi_> \eta_>} \right) + \mbox {Li}_2 \left( e^{ - i \alpha}
\frac{\xi_< \eta_<}{\xi_> \eta_>} \right) \right]\,\, .
\end{equation}
For the $(k = 0$)--term in Eq.\ (\ref{bigj}) one can show that since
\begin{equation}
\int_\mu^\infty \frac{dl}{l} J_0 (l |{\underline x} - {\underline
x}_i|) J_0 (l \tau) = - \ln \left( \frac{\xi_> \mu}{2} \right) -
\gamma\,\,\, ,
\end{equation}
with $\gamma$ being the Euler constant, this term can be
rewritten as 
\begin{equation}
\ln (\xi_> \lambda) \ln (\eta_> \lambda)\,\, ,
\end{equation}
where the cutoff $\lambda$ includes the previous cutoff $\mu$ and all
other numerical prefactors. We finally obtain the answer:
\begin{equation}
{\cal J} = \ln (\xi_> \lambda) \ln (\eta_> \lambda) + \frac{1}{4}
\left[ \mbox {Li}_2 \left( e^{i \alpha} \frac{\xi_< \eta_<}{\xi_>
\eta_>} \right) + \mbox {Li}_2 \left( e^{ - i \alpha} \frac{\xi_<
\eta_<}{\xi_> \eta_>} \right) \right]\,\, .
\end{equation}

\section{}

 The goal of this appendix is to perform the following integration:
\begin{equation}
{\cal I} = \int \frac{d^2 {\underline q}}{{\underline q}^2
(\underline{k}- \underline{q})^2} \left( 1 - \frac{J_1(2|{\underline
q}|a )}{|{\underline q}|a} \right) \left( 1 -
\frac{J_1(2|\underline{k}-\underline{q}|a)}{|\underline{k} -
\underline{q}|a} \right)\,\, .
\end{equation}
Defining $q \equiv |{\underline q}|$ and $p \equiv |\underline{k} -
\underline{q}|$ we rewrite the integral as 
\begin{equation}\label{step1}
{\cal I} = \int \frac{J(p,q) dp \, dq}{q^2 p^2} \left( 1 - \frac{J_1(2
q a )}{q a} \right) \left( 1 - \frac{J_1(2 p a )}{p a} \right)\,\, ,
\end{equation}
where the Jacobian $J(p,q)$ is given by [see \cite{Mueller,ww}]
\begin{equation}\label{jac}
J(p,q) = 2 \pi \, p \, q \, \int_0^{\infty} b \, db \, J_0 (b k) \,
J_0 (b q) \, J_0 (b p)\,\, ,
\end{equation}
with $k = |\underline{k}|$. Inserting Eq.\ (\ref{jac}) into
Eq.\ (\ref{step1}) we obtain:
\begin{equation}\label{step2}
{\cal I} = 2 \pi \, \int_0^{\infty} b \, db \, J_0 (b k) \, \left[
\int_0^{\infty} \frac{dq}{q} J_0(b q) \left( 1 - \frac{J_1(2 q a )}{q
a} \right) \right]^2\,\, .
\end{equation}
The integration over $p$ in Eq.\ (\ref{step1}) is identical to the
integration over $q$, which allowed us to square the $q$--integral
in Eq.\ (\ref{step2}). Performing the integral in the square brackets
we get [see \cite{Mueller,ww}]
\begin{equation}
\int_0^{\infty} \frac{dq}{q} J_0(b q) \left( 1 - \frac{J_1(2 q a )}{q
a} \right) = \theta (2 a - b) \left[ \ln \left(\frac{2 a}{b}\right) - 
\frac{1}{2} + \frac{b^2}{8 a^2} \right]\,\, ,
\end{equation}
which yields for the original integral:
\begin{eqnarray}\label{step3}
{\cal I} & = & 2 \pi \, \int_0^{2 a} b \, db \, J_0 (b k) \, \left[
\ln \left(\frac{2 a}{b}\right) - \frac{1}{2} + \frac{b^2}{8 a^2}
\right]^2 \nonumber\\ & = & 2 \pi \, (2 a)^2 \int_0^1 t \, dt \, J_0 (
2 a k t) \, \left( \ln t + \frac{1}{2} (1 - t^2) \right)^2 \,\, ,
\end{eqnarray}
with $t = b / 2a$.  The integral in Eq.\ (\ref{step3}) can be
calculated by expanding the Bessel function in a power series,
performing the integration in each term, and finally resumming the series.
The result is
\begin{equation}\label{step4}
{\cal I} = 2 \pi a^2 \left( - \frac{1 + (k a)^2 }{(ka)^4} \, \, 
\left( 1 - J_0 (2 k a) \right) + \frac{2 + (k a)^2 }{(ka)^4} \, \, J_2 (2 k a) 
\nonumber \\ +  {_2}F_3 (1,1;2,2,2; - k^2 a^2) \right).
\end{equation}
This expression is used to obtain Eq.\ (\ref{final2}). Taking the $k
\rightarrow 0$ limit of formula (\ref{step4}) we obtain the
integral appearing in the number of elastic scattering events
[Eq.\ (\ref{nelsc})]:
\begin{equation}\label{step5}
\int \frac{d^2 {\underline q}}{({\underline q}^2)^2} \left( 1 -
\frac{J_1(2|{\underline q}|a )}{|{\underline q}|a} \right)^2 =
\frac{5}{6} \pi a^2.
\end{equation}

\end{document}